\documentclass[aps,prl,preprint,superscriptaddress]{revtex4-1}
\pdfoutput=1 

\usepackage{graphicx}
\usepackage{color}
\usepackage{array}
\usepackage{natbib}

\begin{document}

\title{Fabrication of superconducting through-silicon vias}

\author{Justin L. Mallek}
    \thanks{Authors contributed equally to this work}
    \affiliation{MIT Lincoln Laboratory, 244 Wood Street, Lexington, MA 02421}
    \email{justin.mallek@ll.mit.edu}
\author{Donna-Ruth W. Yost}
    \thanks{Authors contributed equally to this work}
    \affiliation{MIT Lincoln Laboratory, 244 Wood Street, Lexington, MA 02421}
\author{Danna Rosenberg}
    \affiliation{MIT Lincoln Laboratory, 244 Wood Street, Lexington, MA 02421}
\author{Jonilyn L. Yoder}
    \affiliation{MIT Lincoln Laboratory, 244 Wood Street, Lexington, MA 02421}
\author{Gregory Calusine}
    \affiliation{MIT Lincoln Laboratory, 244 Wood Street, Lexington, MA 02421}
\author{Matt Cook}
    \affiliation{MIT Lincoln Laboratory, 244 Wood Street, Lexington, MA 02421}
\author{Rabindra Das}
    \affiliation{MIT Lincoln Laboratory, 244 Wood Street, Lexington, MA 02421}
\author{Alexandra Day}
    \affiliation{MIT Lincoln Laboratory, 244 Wood Street, Lexington, MA 02421}
\author{Evan Golden}
    \affiliation{MIT Lincoln Laboratory, 244 Wood Street, Lexington, MA 02421}
\author{David K. Kim}
    \affiliation{MIT Lincoln Laboratory, 244 Wood Street, Lexington, MA 02421}
\author{Jeffery Knecht}
    \affiliation{MIT Lincoln Laboratory, 244 Wood Street, Lexington, MA 02421}
\author{Bethany M. Niedzielski}
    \affiliation{MIT Lincoln Laboratory, 244 Wood Street, Lexington, MA 02421}
\author{Mollie Schwartz}
    \affiliation{MIT Lincoln Laboratory, 244 Wood Street, Lexington, MA 02421}
\author{Arjan Sevi}
    \affiliation{MIT Lincoln Laboratory, 244 Wood Street, Lexington, MA 02421}
\author{Corey Stull}
    \affiliation{MIT Lincoln Laboratory, 244 Wood Street, Lexington, MA 02421}
\author{Wayne Woods}
    \affiliation{MIT Lincoln Laboratory, 244 Wood Street, Lexington, MA 02421}
\author{Andrew J. Kerman}
    \affiliation{MIT Lincoln Laboratory, 244 Wood Street, Lexington, MA 02421}
\author{William D. Oliver}
    \affiliation{MIT Lincoln Laboratory, 244 Wood Street, Lexington, MA 02421}
    \affiliation{Research Laboratory of Electronics, Massachusetts Institute of Technology, 77 Massachusetts Avenue, Cambridge, MA 02139}
    \email{william.oliver@mit.edu}

\date{\today}

\begin{abstract}
Increasing circuit complexity within quantum systems based on superconducting qubits necessitates high connectivity while retaining qubit coherence. Classical micro-electronic systems have addressed interconnect density challenges by using 3D integration with interposers containing through-silicon vias (TSVs), but extending these integration techniques to superconducting quantum systems is challenging. Here, we discuss our approach for realizing high-aspect-ratio superconducting TSVs\textemdash 10 $\mu$m wide by 20 $\mu$m long by 200 $\mu$m deep\textemdash with densities of 100 electrically isolated TSVs per square millimeter. We characterize the DC and microwave performance of superconducting TSVs at cryogenic temperatures and demonstrate superconducting critical currents greater than 20 mA. These high-aspect-ratio, high critical current superconducting TSVs will enable high-density vertical signal routing within  superconducting quantum processors.

\end{abstract}

\maketitle

\section{Introduction}
Superconducting qubits are rapidly advancing from prototype, few-qubit systems to the stage where tens of qubits can be manipulated with error rates low enough to outperform the world's leading supercomputer \cite{Google_QS}. However, further increases in qubit number require the development of scalable architectures capable of addressing many qubits while maintaining coherence. Addressing a 2D array of coupled qubits, such as those needed for most quantum error correction protocols \cite{Fowler_2012}, is prohibitively difficult if qubits are accessed laterally from the periphery of a chip due to interconnect crowding. The difficulty associated with interconnect crowding is significantly eased by utilizing non-planar interconnect geometries. At the most basic level, superconducting air bridges, which arch over signal traces on a single routing layer, \cite{Chen_2014,MicrowaveMag3Dint} and/or flip-chip integration between planar, single-routing-layer chips \cite{Foxen_2018,Obrien_2017,Rosenberg_2017}, bring traces out of plane so they can cross other circuit elements. However, these approaches can result in unwanted spurious couplings due to the proximity of the qubits to the air bridges and/or the flip chip. These limitations can be partially overcome through the use of clever routing schemes, but extending past a few hundred qubits will be challenging. Therefore, while the use of flip-chip integration may carry the field into the onset of the so-called ``noisy intermediate scale quantum" (NISQ) era \cite{NISQ}, it alone is unlikely to enable the next generation of larger-scale NISQ and error-corrected quantum processors.

Fully exploiting the third dimension will require multiple layers of traces for both efficient signal routing and the reduction of interconnect density on the qubit plane. Large-scale classical computing architectures typically employ monolithic fabrication methods to enable high density traces within multiple metallization layers separated by interlayer dielectrics. In some cases, such as for superconducting digital logic applications, the entire metallization stack is superconducting to avoid issues associated with heating from ohmic losses \cite{Tolpygo_2015,hypres}. While monolithic integration of qubits may seemingly resolve the interconnect density problem, these approaches\textemdash as we know them today\textemdash are incompatible with high-coherence qubits due to the lossy interlayer dielectrics \cite{mrsbulletin_2013} which contain defects, generally modeled as two-level systems \cite{AHV_1972, Phillips1972}, that interact with the qubits' electric fields and lower their lifetimes \cite{Schick_1977,Martinis_2005}. Shielding of the qubits by a superconducting ground plane can alleviate some of these issues, but this method is limited by (1) the desire to keep the qubit mode volume large to avoid significant participation with surface defects at interfaces and native oxides \cite{Martinis_2005,Woods_2019}, and (2) the need to bring signals out of the multilayer stack near the qubits. Another option could be to create a large separation between the qubit chip and the monolithic multilayer chip, but such a large separation necessitates the use of large-diameter bumps to achieve a large chip-to-chip stand-off, limiting the circuit density.

\begin{figure*}[htb]
\includegraphics[width = 6 in]{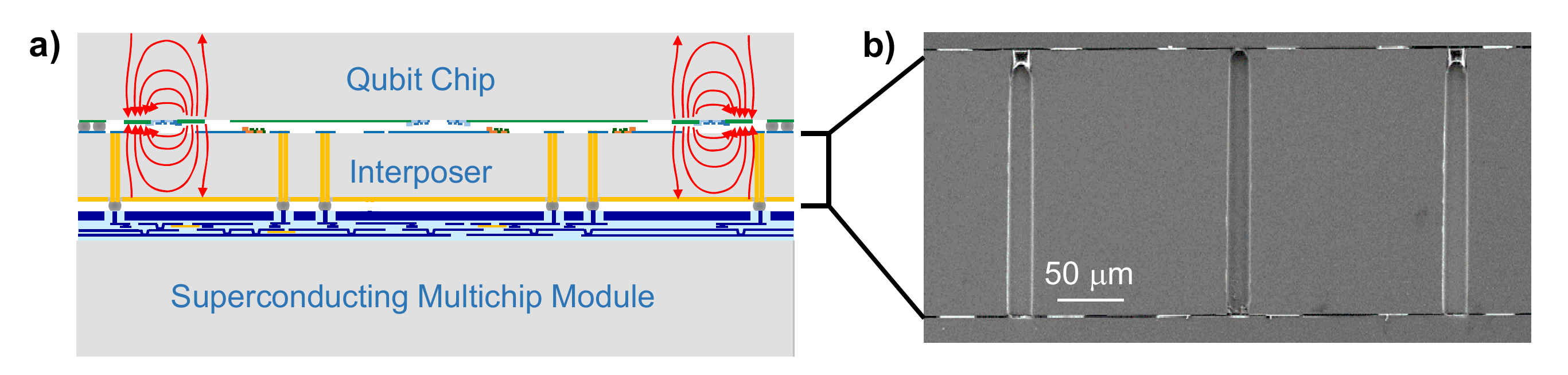}%
\caption{\label{fig:ThreeStack}  a) Schematic cross section of the three-tier quantum processor depicting the locations of superconducting qubits, superconducting TSVs and the superconducting multichip module (SMCM). Signals can travel from the SMCM chip through indium bumps (dark gray) connecting the top metallization layer of the SMCM to the bottom of the interposer chip, through the TSVs, and to the top of the interposer chip. From the top of the interposer chip, circuit elements can couple to qubits capacitively and inductively across the gap or galvanically through another bump to the qubit chip. b) Scanning electron microscope (SEM) cross section of a double-bump-bonded three-tier stack, which was mechanically polished from the side to reveal the TSVs and indium bump bonds. Underfill adhesive, visibly extruding into the top of the TSVs, was added to increase mechanical integrity during polishing and is not present in a functional device.
 }
\end{figure*}

We have previously described an alternate approach that enables the 3D integration of superconducting qubits with multilayer superconducting circuits \cite{Tolpygo_2015,hypres} while retaining qubit coherence \cite{Rosenberg_2017,Yost2020}. Our approach relies on the development of an interposer chip with superconducting through-silicon vias (TSVs). As shown in Figure \ref{fig:ThreeStack}, this chip is integrated within a three-tier quantum processor comprised of a chip with high coherence superconducting qubits on the top tier and a superconducting multi-chip module (SMCM) on the bottom tier. The superconducting TSVs connect metallization layers on the top and the bottom of the interposer chip and can thereby deliver signals vertically from the SMCM to the qubit chip without exposing the qubits to the lossy dielectrics of the SMCM. In addition, the superconducting TSVs enable new functionality that utilizes the third dimension, for example, to selectively couple readout resonators and microwave drive lines to qubits, or within the qubit circuit itself as part of the qubit capacitor \cite{Schwartz_tbp}. Superconducting TSVs can also reduce spurious coupling by controlling the electric field in substrates near the qubits, for example, by moving spurious box modes in silicon chips to higher frequencies to reduce their impact on qubits. 

Superconducting TSVs that enable extensible qubit circuits must fulfill several basic requirements. First, they must have a small footprint in order to accommodate a high density of interconnects. Current state-of-the-art gate-based superconducting qubits are a few hundred micrometers on a side, resulting in a qubit density of approximately 10/mm$^2$. To afford multiple, shielded control and readout interconnects per qubit without reducing qubit density, the TSVs should be of order $10~\mu $m on a side or smaller. Another requirement is that the interposer chip be thick, 100 $\mu $m or more, to sufficiently space the qubit chip from the SMCM and to allow for handling and bonding of the interposer chip. Together, the high-density and thick-interposer requirements translate to small-diameter TSVs with high-aspect ratios. Since the superconducting TSVs will be used to carry signals to the qubit chip, it is also important that the material in the TSVs has a critical temperature much larger than the qubit operating temperature, typically less than 100 mK. Finally, the superconducting critical current density must be high enough to support a single TSV critical current of $\sim 10~m$A to enable flux biasing of the tunable qubits in common use today. 

Fabricating superconducting TSVs with high critical temperatures and high critical currents is challenging due to the high-aspect-ratio geometry imposed by our design requirements. Previous work in the field has focused on TSVs that are either very large with low-aspect ratios \cite{Vahidpour_2017} or have low critical currents \cite{Jhabvala_2014}. In this paper, we describe the development and characterization of high-aspect-ratio, high-yield superconducting TSVs that are suitable for use in a quantum processor based on superconducting qubits.

\section{Interposer fabrication}

\begin{figure*}[ht]
\includegraphics[width = 6.5 in]{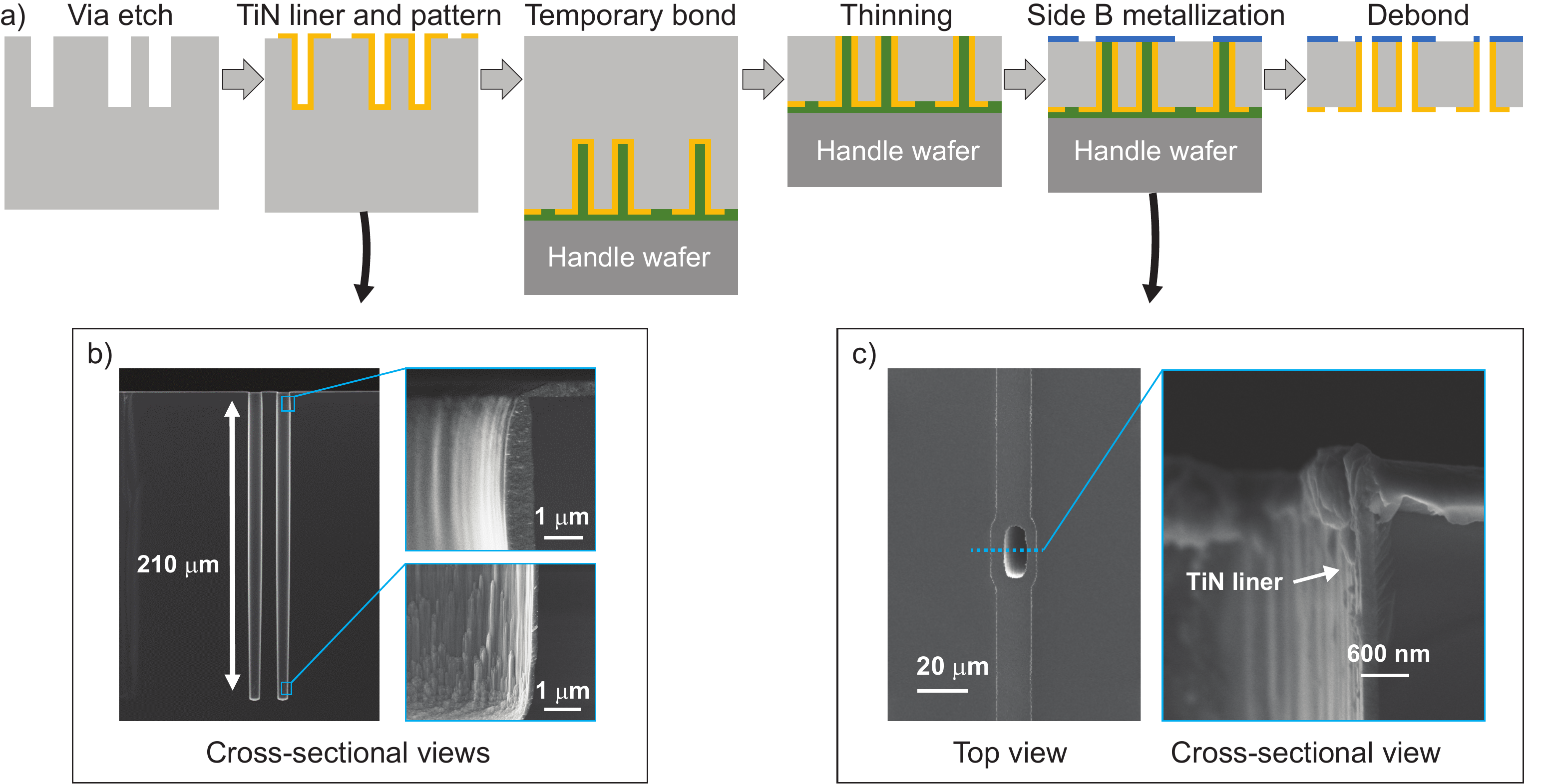}%
\caption{\label{fig:Fabflow}  a) Schematic depiction of the interposer fabrication process. Fabrication began with dry etching blind vias in side A of the wafer, followed by CVD deposition of the superconducting TiN liner. Next, side A was temporary bonded to a handle wafer and the interposer wafer was thinned from side B, revealing the TSVs. Side B metallization completed the device structures and the devices were debonded from the handle wafer. b) SEM cross-section showing the 210 $\mu$m deep vias, and insets of the TiN liner at the opening and base of a via. c) Top-down and cross-sectional SEM images of side B of the device wafer after metallization. In the top view, the aluminum trace is visible running vertically and surrounding the TSV.}
\end{figure*}

The TSV interposer fabrication sequence is shown in Figure \ref{fig:Fabflow}. We utilize a via-first fabrication process which allows thermally sensitive Josephson junctions to be fabricated after the completion of high-temperature processing steps. Each 200-mm diameter, 725-$\mu$m thick, high-resistivity silicon wafer received a standard RCA clean to remove trace surface contaminants. A silicon dioxide hardmask is deposited onto the front (side A) of the wafer, and the via openings are lithographically defined and dry etched into the hardmask. We then etch high-aspect-ratio blind vias approximately 210 $\mu$m deep into the wafer using a deep silicon etch system. The via sidewalls are smoothed of minor scalloping by growing a sacrificial silicon dioxide liner and stripping the oxide liner using buffered oxide etch (BOE). The final via opening dimensions are 10x24 $\mu$m$^{2}$.

After the vias are prepared, they are lined with superconducting material. This is challenging as the via aspect ratio is high ($>$10:1) and the sidewalls are nearly vertical, approximately 89.7$^{\circ}$. After consideration of multiple deposition methods\textemdash including electroplating, physical vapor deposition (PVD), atomic layer deposition (ALD), and chemical vapor deposition (CVD)\textemdash we developed a CVD-based process for depositing superconducting titanium nitride (TiN) with high conformality. Typical TiN film thicknesses are 360 nm at the surface, 165 nm at a via depth of 100 $\mu$m, and 150 nm at the base of the 210-$\mu$m-deep via. The TiN on the surface of side A is then lithographically patterned and dry etched to form via interconnects and traces. 

Interposer fabrication continues by temporary wafer bonding side A of the interposer wafer to a silicon handle wafer. Temporary bonding protects side A from wafer handling damage and debris, and it provides structural support during wafer thinning and subsequent fabrication steps on the thinned wafer. The vias are revealed using a combination of grinding and chemical mechanical polishing (CMP), reducing the thickness of the interposer wafer to 200 $\mu$m; we now refer to this surface as side B. Traces and interconnects are formed on side B by evaporated aluminum deposition and liftoff. 

Wafer-scale room temperature electrical measurements are completed while the interposer wafer remained temporary bonded to the handle wafer. Cryogenic measurements of superconducting critical temperature ($T_c$) and critical current ($I_c$) are made on diced chips where the bonded wafer pair remained temporary bonded during the measurements. Resonator measurements are made on diced chips which were debonded prior to packaging.

\section{Room temperature and cryogenic electrical measurements of TSV structures}

\begin{figure*}[ht]
\includegraphics[width = 6 in]{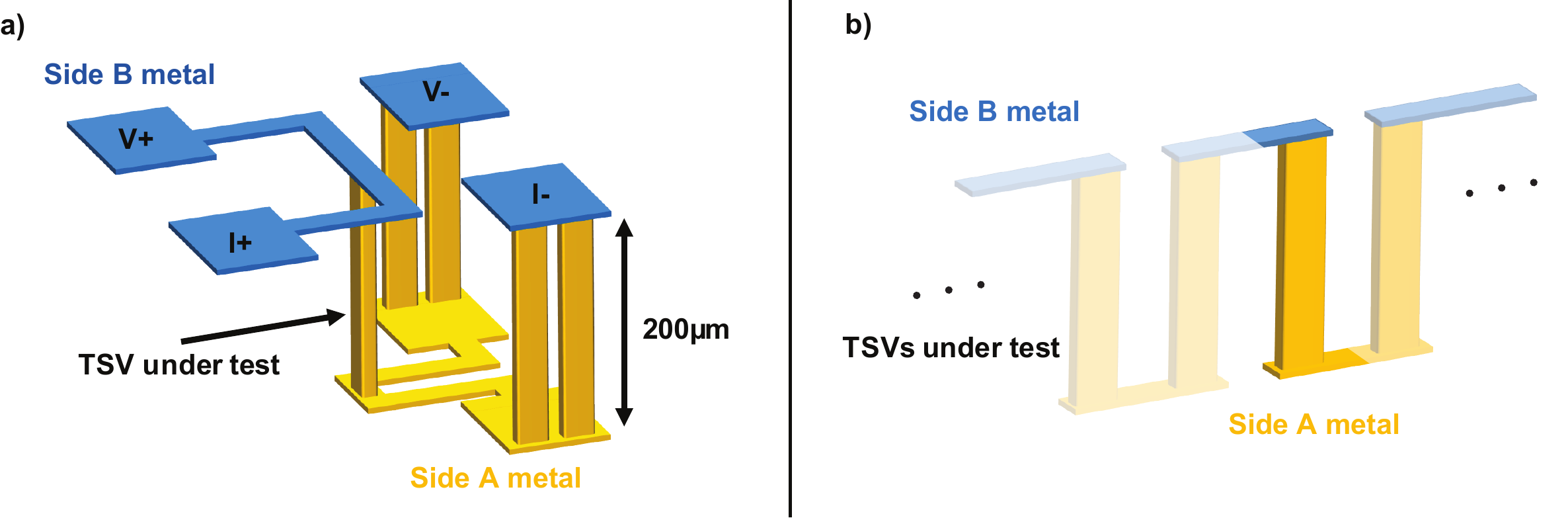}%
\caption{\label{fig:CBKR_chains}  Electrical test structures used for room temperature and cryogenic TSV measurements.  a) Cross Bridge Kelvin Resistor (CBKR) test structure with four-point probe pads on side B of the wafer. The central TSV is under test and the TSVs within the pads provide electrical continuity from side A to side B. b) Section of a TSV chain structure where up to 400 superconducting TSVs in series form the test circuit.}
\end{figure*}

We evaluated the performance of both individual TSVs and serial chains of TSVs using room temperature and cryogenic measurements. As shown in Figure \ref{fig:CBKR_chains}a, a cross-bridge Kelvin resistor (CBKR) \cite{CBKR} is used to test individual TSVs. This four-wire CBKR structure was designed so that key metrics of TSV performance\textemdash room temperature resistance, superconducting critical temperature, and superconducting critical current\textemdash are dominated by the TSV liner with negligible contributions from the planar wiring or ancillary TSVs. Chain structures, represented in Figure \ref{fig:CBKR_chains}b, contain 400 TSVs connected in series by superconducting traces on sides A and B. Each "link" in the chain contains a single TSV and a length of the superconducting trace on sides A and B which is half the distance to the adjoining TSVs. This chain structure, where even a single faulty TSV would cause the entire chain to fail, is useful for evaluating TSV yield in cases where the yield is intrinsically high.

\begin{figure*}[ht]
\includegraphics[width = 6 in]{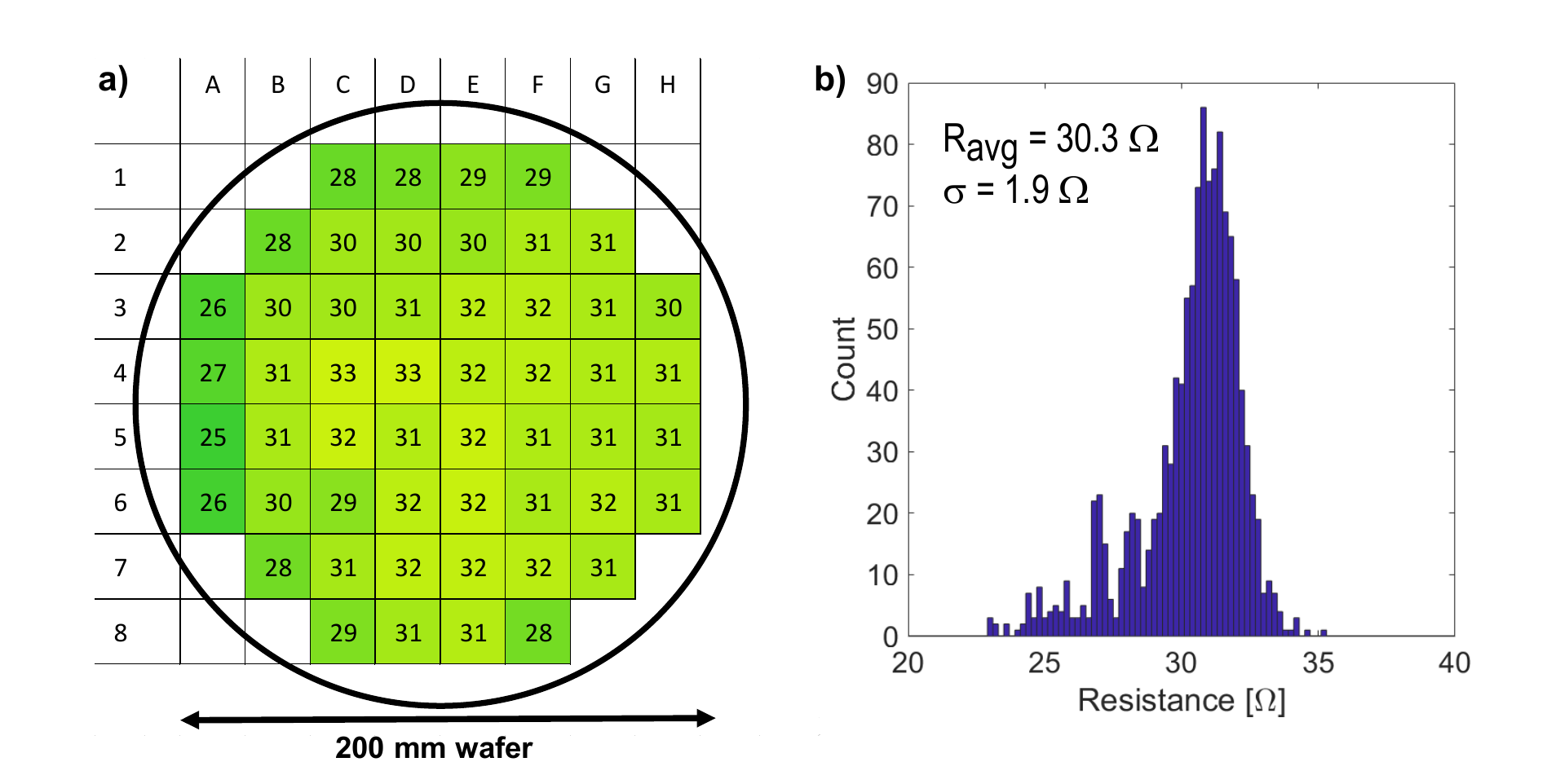}%
\caption{\label{fig:RT_res}  Results of room temperature electrical testing. a) A wafer map showing average single-TSV CBKR resistances (from 24 measurements per reticle) across 52 reticles.  b) Histogram of the full wafer TSV resistance measurements. }
\end{figure*}

Room temperature measurements are used for rapid feedback on the fabrication process and to gauge cross-wafer uniformity and yield. A semi-automated wafer prober is used to measure the resistance of 1,248 single-TSV CBKRs (Figure \ref{fig:CBKR_chains}a) and 312 chains of 400 TSV links (Figure \ref{fig:CBKR_chains}b) across each wafer. Figure \ref{fig:RT_res} shows results of the CBKR resistance measurements. The CBKR resistance across the wafer is highly uniform with only a slight radial dependence attributed to (1) a radial dependence of the CMP polish rate, where the edge polishes slightly faster, and (2) a slight asymmetry along the left edge of the wafer (column A) which is attributed a nonuniformity of gas flow within the CVD tool process chamber. The full-wafer average resistance of the CBKR structures is 30.3 $\Omega$ with a standard deviation of 1.9 $\Omega$, as shown in Figure \ref{fig:CBKR_chains}b. Of the 312 400-link chains that were measured, we observed four outliers with a resistance of an order of magnitude higher than the resistance of rest of the chains. After these data points were removed, the TSV chains for this wafer have an average per-link room temperature resistance of 36.8 $\Omega$ with a standard deviation of 1.7 $\Omega$. This is consistent with the measured resistance of individual TSVs and the additional resistance contributed by the metallization on sides A and B that formed each link.

\begin{figure*}[ht]
\includegraphics[width = 4 in]{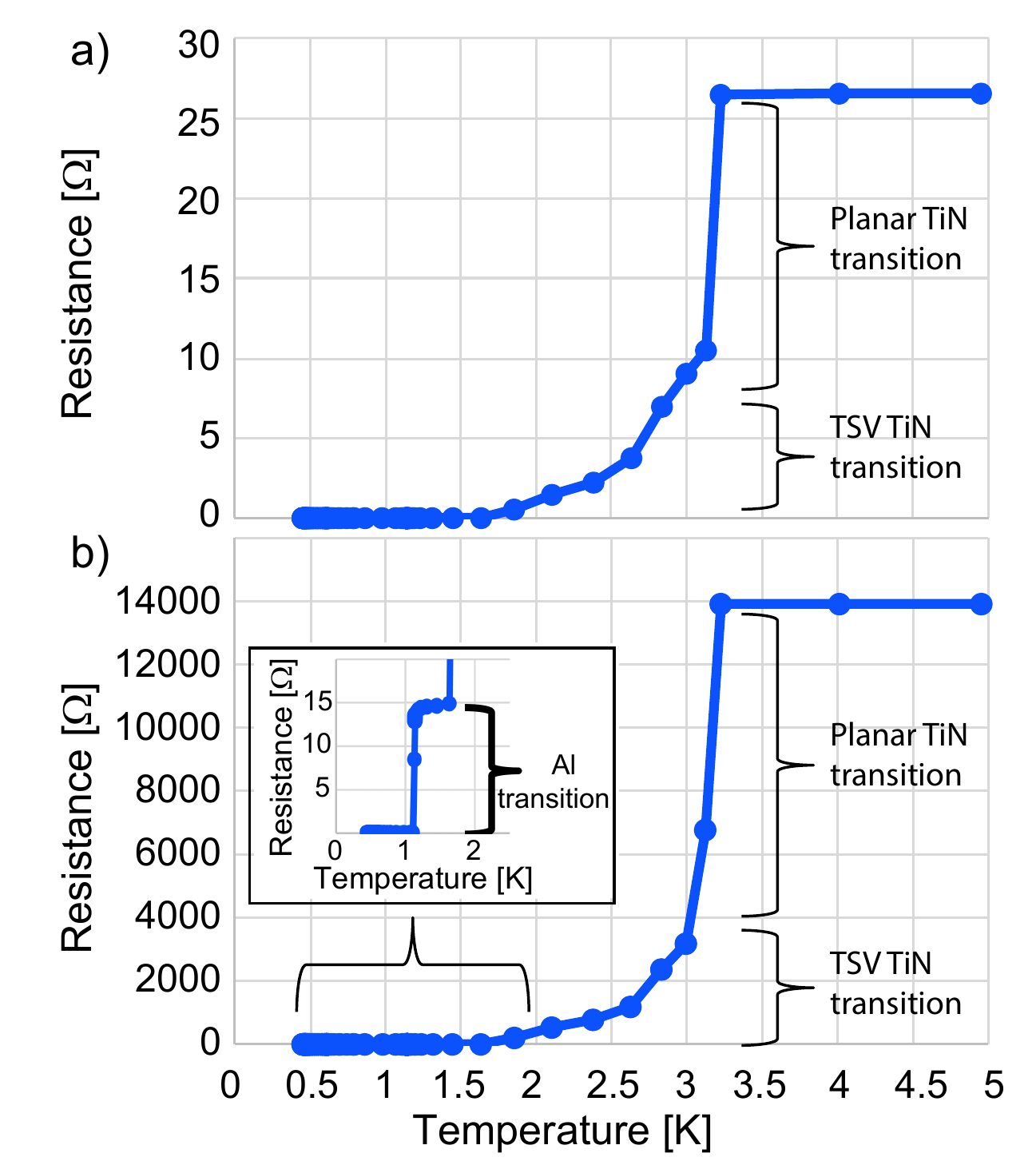}%
\caption{\label{fig:Tc}  a) Superconducting transition of a CBKR test structure where a single TSV is under test. b) Superconducting transitions within a 400-link TSV chain for the TiN metallization on side A, the TSV TiN liner, and the Al metallization on side B (inset).}
\end{figure*}

Cryogenic measurements are used to qualify TSV superconducting performance. To measure the $T_c$ and $I_c$ of the TSVs, diced chips with CBKR and chain test structures are packaged and mounted onto the base-temperature stage of a dilution refrigerator and a thermometer is mounted to the package to measure temperature. We perform $T_c$ measurements by monitoring the resistance and temperature of the test structures while changing the temperature of the refrigerator. To confirm the package is well thermalization, $T_c$ measurements were made while warming and cooling the refridgerator and the superconducting transitions were found to be well-matched. Figures \ref{fig:Tc}a and \ref{fig:Tc}b show $T_c$ data for a single TSV and for a chain of 400 links, respectively. When measuring the critical temperature of the TSV chain, there is an initial drop in resistance from approximately 14 k$\Omega$ at 3.2 K down to 15 $\Omega$ at 1.6 K. At this temperature the TiN lining of the TSVs, as well as the planar TiN deposited on side A of the wafer in the same fabrication step, is superconducting. The remaining 15 $\Omega$ at 1.6 K is attributed to the aluminum metal connecting the TSV chain together on side B. At 1.1 K the aluminum also becomes superconducting and the resistance drops close to $0~\Omega$. A 4-wire measurement using a lock-in amplifier was performed to place an upper bound of $100~n\Omega$ per link, limited by the noise sensitivity of the measurement related to the common-mode rejection ratio of the amplifier and thermal voltage offsets.

We perform $I_c$ measurements by monitoring the resistance of the test structure while slowly sweeping the current from zero to the point where the resistance increased sharply as the metal transitioned from the superconducting to the normal state. These measurements are performed near the base temperature of the refrigerator, much less than the $T_c$ of the titanium nitride film ($\sim 3~K)$ and aluminum film ($\sim 1.1~K$). Four individual CBKR structures were measured and found to have an average $I_c$ of 22 mA with a standard deviation of 2.2 mA. Additionally 5 400-link chains (2000 TSVs in total), were measured and have an average $I_c$ of 14 mA with a standard deviation of 1.0 mA.

\section{TSV-interrupted resonators}
\begin{figure}[ht]
\includegraphics[width = 6 in]{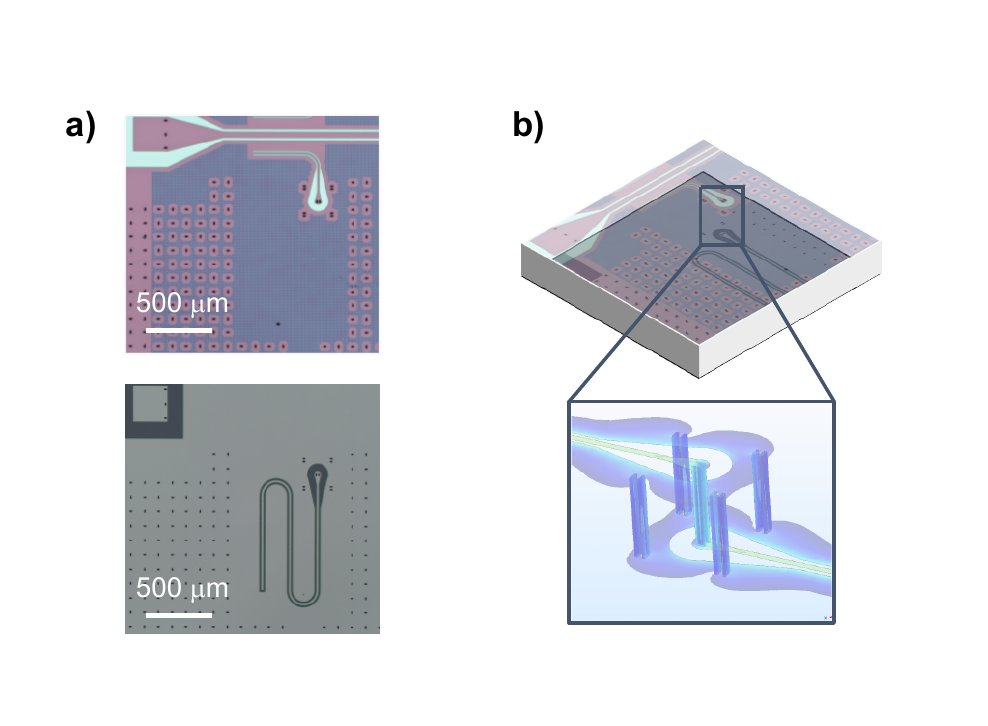}
\caption{\label{fig:TSVres}  Quarter-wave coplanar waveguide transmission line resonators with TSV transitions. a) Optical images of side A with patterned TiN (top) and side B with patterned Al (bottom). The bottom image has been mirrored over the x axis for clarity. b) 3D schematic of the resonator. The inset shows an electromagnetic simulation of the current flow near the TSV transition. A pair of TSVs in parallel connects the center trace of the coplanar waveguide, and four pairs of TSVs connect the ground plane on either side of the chip.}
\end{figure}

In addition to providing direct current (DC) signals to qubits, TSVs may be used within a quantum processor to provide microwave signals to the qubits, or even as part of a structure used to store quantum information. To test the RF performance, we embedded TSVs in superconducting transmission line resonators and measured their quality factors. As shown in Figure {\ref{fig:TSVres}}a, the resonators comprised two or more sections of a coplanar waveguide quarter wave resonator, split between sides A and B of a chip with TSVs. Simulations indicate that a single TSV would introduce a significant impedance mismatch at microwave frequencies, so we designed a TSV transition from side A to side B that included two TSVs connecting the center traces of the transmission lines and eight TSVs connecting the ground planes, as shown in {\ref{fig:TSVres}}b. The shape of the transition was adjusted to minimize reflections at the TSV interface. 

\begin{figure}[ht]
\includegraphics[width = 4 in]{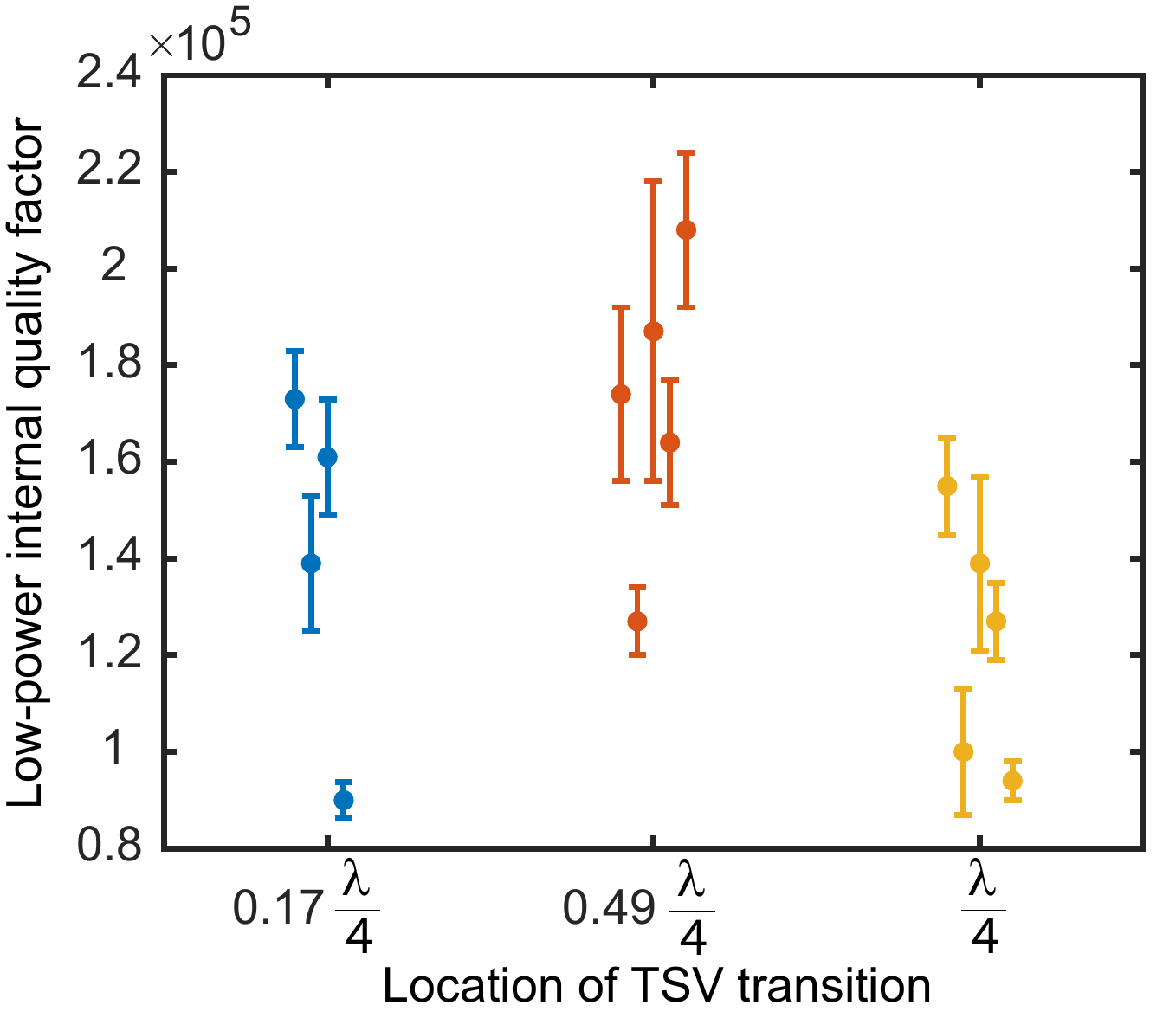}
\caption{\label{fig:TSVresQ}  Single-photon internal quality factor of quarter wave coplanar waveguide resonators with TSV transitions as a function of location of the transition, as measured from the shorted end of the transmission line. Note that the location of the TSV transition for each resonator matches the tick labels on the x axis, and the data points have been artificially spread out for clarity.}
\end{figure}

By placing the TSV transition at different locations within the resonator, we can determine the type of loss\textemdash electric or magnetic\textemdash that is potentially introduced by the TSVs. If the transition is placed at the voltage antinode, it will be maximally sensitive to loss arising from two-level system defects, which limit the quality factor of resonators by interacting with the electric field. If the transition is placed at the current antinode, it will be maximally sensitive to resistive losses. Figure \ref{fig:TSVresQ} shows measurements of the quality factor for TSV-interrupted resonators of the type shown in Figure \ref{fig:TSVres} for three different positions of the TSV transition, ranging from close to the voltage antinode (blue data points) to the current antinode (yellow data points). The data show no clear trend with the position of the TSV transition, and the data are consistent with measured quality factors of planar resonators on each of side A and side B, showing the TSVs do not introduce significant microwave loss.

\section{Summary}
We have successfully fabricated high-aspect-ratio superconducting TSVs on a thinned 200-mm silicon wafer using a via-first fabrication process, conformal CVD titanium nitride process, and back-side wafer thinning. Room temperature electrical measurements indicate high intra-wafer uniformity of the resistance measured through the TSVs. After dicing individual test chips, measurements are conducted in a dilution refrigerator to evaluate the critical temperature and critical current of the superconducting TSVs; critical temperatures as high as 1.6 K and critical currents as high as 22 mA were measured. We evaluated resonators with embedded TSVs transitions at various positions within the resonator trace and measured quality factors consistent with planar resonators fabricated on each side of the thinned wafer. The high-aspect-ratio superconducting TSVs demonstrated here enable high-density, vertical signal routing within a superconducting quantum processors without negatively impacting current levels of qubit performance.

\section{Acknowledgements}
We gratefully acknowledge the MIT Lincoln Laboratory design, fabrication, packaging, and testing personnel for valuable technical assistance. 
This research was funded in part by the Office of the Director of National Intelligence (ODNI), Intelligence Advanced Research Projects Activity (IARPA) and the Defense Advanced Research Projects Agency (DARPA) under Air Force Contract No. FA8702-15-D-0001. The views and conclusions contained herein are those of the authors and should not be interpreted as necessarily representing the official policies or endorsements, either expressed or implied, of the ODNI, IARPA, DARPA, or the U.S. Government.


\bibliographystyle{naturemag}
\bibliography{scqubits}

\end{document}